\documentclass[fleqn,usenatbib, useAMS]{mnras}

\usepackage[T1]{fontenc}
\usepackage{ae,aecompl}
\usepackage[utf8]{inputenc}
\usepackage{threeparttable}
\usepackage{booktabs, caption, makecell}
\usepackage{pdflscape}
\usepackage{geometry}
\usepackage{float}

\usepackage{graphicx}	
\usepackage{amsmath}	
\usepackage{amssymb}	

\usepackage{rotating}
\usepackage{float}
\usepackage{tabularx}
\usepackage{natbib}
\usepackage{hyperref}
\usepackage{svg}
\newcommand{\orcid}[1]{\href{https://orcid.org/#1}{\includegraphics[width=10pt]{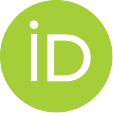}}}
\newcommand{\Teff}{\mbox{$T_{\mathrm{eff}}$}}
\newcommand{\logg}{\mbox{$\log g$}}
\newcommand{\Msun}{\mbox{$\mathrm{M_\odot}$}}

\title[Ly\,$\alpha$ wing opacities in cool white dwarfs]{Re-evaluating Lyman $\alpha$ wing opacities and the low mass-problem in cool white dwarfs}
\author[Sahu et al.]
{Snehalata Sahu\orcid{0000-0002-0801-8745}$^{1}$\thanks{E-mail: snehalatash30@gmail.com}, 
Pier-Emmanuel Tremblay\orcid{0000-0001-9873-0121}$^{1}$, Detlev Koester\orcid{0000-0002-6164-6978}$^{2}$, Mairi W. O'Brien\orcid{0000-0003-0928-4065}$^{1}$, \newauthor{Simon Blouin\orcid{0000-0002-9632-1436}$^{3}$, Boris T. G\"ansicke\orcid{0000-0002-2761-3005}$^{1}$, Vince Fairchild$^{1}$}\\\\
$^{1}$ Department of Physics, University of Warwick, Coventry, CV4 7AL, UK\\
$^{2}$ Institut f\"ur Theoretische Physik und Astrophysik, University of Kiel, D-24098 Kiel, Germany\\
$^{3}$ University of Victoria, Department of Physics and Astronomy, Victoria, BC V8W 2Y2, Canada\\
}
\date{Accepted 2025 October 31; Received 2025 October 28; in original form 2025 July 22}
\begin{document}
\label{firstpage}
\pagerange{\pageref{firstpage}--\pageref{lastpage}}
\maketitle

\begin{abstract}
\textit{Gaia} observations have reignited interest in the optical and ultraviolet (UV) opacity problems of cool white dwarfs ($\Teff\leq6000$\,K), which were thought to be resolved nearly two decades ago through the inclusion of Lyman\,$\alpha$ red wing opacity arising from H-H$_2$ collisions in atmospheric models. Recent studies have revealed that their masses derived from \textit{Gaia} optical photometry are 0.1$-$0.2\,\Msun\ lower than expected from single-star evolution. Since the Ly \,$\alpha$ H-H$_2$ wing opacity significantly affects the blue end of their optical spectra, it may contribute to the mass discrepancy. To investigate this hypothesis, we revisited the Ly\,$\alpha$ opacity calculations in the quasi-static single and multi-perturber approximations by explicitly using the \textit{ab initio} potential energy data of H$_3$ while fully accounting for the H-H$_2$ collision angle. We find that the opacity is slightly smaller than the standard models at the shortest wavelengths ($\leq5000$\,\AA), but larger at longer wavelengths. Comparing synthetic magnitudes (\textit{GALEX, Gaia, WISE}) to the observations of the 40\,pc white dwarf sample, we note that the revised models tentatively reproduce the observed $NUV-G$ colours for stars cooler than 6000\,K, but still fail to match $G_{\rm BP} - G_{\rm RP}$ colours, resulting in similarly low inferred masses ($\leq 0.5$\,\Msun) as obtained with the standard Ly\,$\alpha$ opacity. Exploring other dominant opacity sources, we discover that decreasing the strength of the bound-free H$^-$ opacity in existing models better reproduces the optical and infrared colours, while collision-induced absorption (CIA) opacity is ineffective in resolving the low-mass problem. We highlight the need for improved opacities and multi-wavelength observations in future studies.  
\end{abstract}


\section{Introduction}
One of the major challenges in white dwarf astrophysics is to accurately model gas opacities in cool objects having effective temperatures below $\approx$6000\,K \citep{saumon2022}, in order to predict their spectral energy distributions. 
In this work, we focus on the pure-hydrogen atmospheres, which constitute $\approx$75\,per cent of cool white dwarfs \citep{blouin2019,Elms2022, Brien2024, Kilic2025}.
Three main opacities operate in the high-density regime of these white dwarf atmospheres ($\rho\gtrsim 10^{-5} $\,g\,cm$^{-3}$), which include bound-free H$^-$ absorption in the optical \citep{john1988}, Ly\,$\alpha$ red wing opacity in the optical blue/near-UV arising due to the collisions of H atoms with H$_2$ molecules \citep{kowalski2006, Rohrmann2011}, and collision-induced absorption (CIA) due to H$_2$-H$_2$ collisions in the infrared \citep{borysow2001}. The refinement of these opacities is crucial for a better determination of the parameters (mass, radius, age) of cool white dwarfs. This, in turn, can have implications on the luminosity function of very old white dwarfs \citep{Fontaine2001, Hansen2002, GO2016, Kilic2017, Roberts2025}, and therefore their application as cosmo-chronometers for probing the age and star formation history of the Galaxy. Furthermore, accurate age and mass estimates are crucial for correctly interpreting old white dwarf planets and planetary systems (for instance, planetary migration, \citealt{diego2020}; origin of lithium, \citealt{kaiser2025}). 

While analysing the photometric observations of 110 cool white dwarfs, \cite{Bergeron1997} reported a flux deficiency in the $B$ band for stars cooler than 6000\,K. They suggested that the pseudo-continuum bound-free opacity from H atoms that is perturbed by collisions with other particles in a dense medium, might be responsible for a missing blue/UV opacity ($\lambda\lesssim5000$\,\AA). However, using realistic H$_2$ and H$_3$ potential curves (which describe the interaction energies of the H-H and H-H$_2$ systems, respectively), \cite{kowalksi2006b} found that this opacity source is too weak to explain the missing blue opacity in the models, therefore ruling it out as a possible solution. Rather, \cite{kowalski2006} (hereafter KS06) attributed the origin of this observed blue flux deficiency to the pressure broadening of the Ly\,$\alpha$ line caused by the collisions of H atoms with H$_2$ molecules. Incorporating their contribution in the model opacities using the quasi-static, single-perturber broadening approximation appropriate for close collisions, they were successfully able to reproduce the observed optical spectra of cool white dwarfs. Later, these results were confirmed by an independent study using a similar approach \citep{Rohrmann2011}. 
Furthermore, \cite{saumon2014} tested the KS06 opacity with \textit{Hubble Space Telescope} (\textit{HST}) STIS spectra of eight cool white dwarfs (4500--5500\,K), and found that the red wing of the Ly\,$\alpha$ line reproduces the rapidly decreasing near-UV flux very well. However, they noticed that the model fits become worse at lower \Teff\ and higher surface gravity (i.e. at higher photospheric densities and pressures), thus highlighting the need for more accurate H-H$_2$ potentials and dipoles, a better quantum collision theory, or a more detailed treatment of non-ideal effects.

The \textit{Gaia} spacecraft brought a revolution by providing precise photometric and parallax measurements of billions of point sources, of which 359\,000 high-confidence white dwarfs were identified using Early Data Release\,3 \citep{nicola2021}. This large sample led to the accurate construction of the field white dwarf Hertzsprung-Russell diagram (HRD), detecting objects as cool as $\approx$3500\,K, and henceforth revealing many interesting features that were never observed before. This includes the discovery of the Q-branch arising from the cooling delay due to crystallization and $^{22}$Ne distillation of C-O core white dwarfs \citep{Tremblay2019, Cheng2019, Bedard2024}, which largely comprises high-mass white dwarfs but extends to cool ($\leq$6000\,K) canonical mass ($\approx$$0.6\,$M$_{\odot}$) remnants. \textit{Gaia} photometric and astrometric measurements led to the detailed analysis of nearby cool white dwarfs \citep{hollands2018,bergeron2019,blouin2019,mccleery2020, Elms2022, Caron2023, Brien2024, Kilic2025}. All studies have so far shown that the white dwarf colours and absolute magnitudes predicted from model atmospheres and evolutionary tracks deviate from the observed \textit{Gaia} $G$ vs $G_{\rm BP}-G_{\rm RP}$ HRD track of cool ($<$6000\,K) DA and DC white dwarfs. When fitting the pure-hydrogen model atmospheres and standard $M-R$ relations to \textit{Gaia} photometry and astrometry, this leads to low inferred masses ($<$0.5\,\Msun) for the coolest white dwarfs ($\Teff\lesssim 5000$\,K). A very similar issue is observed with other optical photometry, such as Pan-STARRS \citep{mccleery2020}. This is at odds with expectations, as white dwarfs are thought to cool at constant mass \citep{Tremblay2016}, and the typical white dwarf mass is around 0.6\,\Msun. This suggests that, despite significant theoretical progress, the issue of optical opacity in cool white dwarfs first noted by \citet{Bergeron1997} may not be fully resolved, especially in light of the more precise and accurate data now available from \textit{Gaia}. To address this issue, \cite{Brien2024} conducted an experiment by changing the strength of the KS06 Ly\,$\alpha$ H-H$_2$ opacity. Multiplying the existing opacity by an ad-hoc factor of five yielded a better match with the observed colours and absolute magnitudes in the \textit{Gaia} HRD, but resulted in greater discrepancies in the UV-optical HRD. This suggests that a simple and not physically motivated multiplication factor is not fully effective in resolving the low-mass issue, thus demanding a thorough investigation.

With this objective in mind, we revisited the Ly\,$\alpha$ wing opacity arising from H-H$_2$ interactions in this work. Note that Ly\,$\alpha$ line profile calculations rely on accurate H$_3$ potential energy surfaces that are derived from quantum chemical methods. The previous KS06 model was based on \cite{boothroyd1996}, considering the interaction energies at an H-H$_2$ angular separation of 90$^\circ$. Here, we explore the effect of the choice of H-H$_2$ potential energy curves in the opacity and model atmosphere calculations, considering various collision angles in Sec.\,\ref{sec:theory}. We present the new model opacities, both using the single- and multiple-perturber collision approximations, with their application to the 40\,pc white dwarf sample in Sec.\,\ref{sec:res}. We discuss their implications for the low mass-problem of cool white dwarfs in Sec.\,\ref{sec:discuss}. 

\section{Theoretical approach}\label{sec:theory}
\subsection{Line broadening theory}
A semi-classical quasi-static approximation is used to calculate the wing opacity of the Ly\,$\alpha$ line following the methodology of KS06. This approach is based on the Frank-Cordon principle, which states that the nuclear separation remains fixed during the electronic transitions of the molecules as a result of a large mass difference between the nucleus and the absorbing electron.

The differential probability d$P$ of finding an H-H$_{2}$ collision pair with an inter-particle separation between $r_{c}$ and $r_{c}$+ d$r_{c}$ in a low-density medium ($<$1 g\,cm$^{-3}$) is
\begin{align}
    {\rm d}P(r_{c}) = n_{\rm pert}\,r_{c}^2\,{\rm d}r_{c} \int_{\theta,\phi}e^{(-E_{1}(r_c,\theta,\phi)/k_{\rm B}T)} \sin\theta\ {\rm d}\theta\ {\rm d}\phi ,\label{eqn:1}
\end{align}
where $n_{\rm pert}$ is the number density of perturbers (H$_2$ in this case), $E_{1}$ is the lower bound state interaction energy between the H atom and the H$_2$ molecule, $\theta$ is the collision angle between the centre of mass of the H$_{2}$ molecule with the H atom, and $k_{\rm B}$ is the Boltzmann constant.

For H-H$_2$ colliding pairs, the Ly\,$\alpha$ transition energy is given by the difference between the ground state and the lowest Rydberg states of the H$_3$ system 
\begin{align}
    &E_{j}(r_c)-E_{1}(r_c)=h\nu\\
    &\frac{{\rm d}r_c}{{\rm d}\nu}= \frac{h} {{\rm d}(E_{j}-E_{1})/{\rm d}r_c} ,\label{eqn:3}
\end{align}
where $\nu$ is the frequency of the absorbed photon.

The line profile $\alpha_{\nu}$ is calculated following a one-perturber quasi-static approximation applicable to the far red wings of the Ly\,$\alpha$ line \citep{Allard1982}:
\begin{align}
    &\alpha_{\nu}{{\rm d}\nu} \sim \sum_{j} {\rm d}P(r_{c}(\nu)) \frac{|T_{1j}|^2}{|T_{12}^0|^2} , \label{eqn:4}
\end{align}
where $T_{1j}$ is the dipole moment corresponding to the transition between the ground state ($E_{1}$) and first Rydberg states ($E_{j}$) of the H$_{3}$ system and $T_{12}^{0}$ is the transition dipole moment of the isolated H atom for Ly\,$\alpha$. Here, the line profile is normalised such that 
\begin{align}
    \int_{0}^{\infty} \alpha_{\nu}{{\rm d}\nu}= 1.
\end{align}
The absorption cross-section is given by
\begin{align}
    &\sigma_{\nu} ({\rm cm}^{2})=\frac{\pi e^{2} f_{12} \alpha_\nu}{m_e c},
\end{align}
where $f_{12}$ is the oscillator strength for the Ly\,$\alpha$ transition.
Substituting for $\alpha_{\nu}$ from Eqn.\,\ref{eqn:4}, we get
\begin{align}
     &\sigma_{\nu}= \frac{\pi e^{2}}{m_e c} \frac{{\rm d}P(r_c)}{{\rm d}r_c} \frac{{\rm d}r_c}{{\rm d}\nu} \sum_{j} \frac{|T_{1j}|^2}{|T_{12}^0|^2}\label{eqn:7}.
\end{align}
Substituting for $({\rm d}P/{\rm d}r)$ and $({\rm d}r_c/{\rm d}\nu)$ from Eqns.\,\ref{eqn:1} and \ref{eqn:3}, we get an expression for the absorption cross-section per unit number density of perturbers $(\sigma_{\nu}/n_{\rm H_{2}}$ in units of ${\rm cm}^{5})$ which integrates to 1 in frequency. 

Finally, the opacity ($\kappa_\nu$) is correlated with the absorption cross-section as
\begin{align}
    \kappa_\nu ({\rm cm^2/g})=\frac {n_{\rm H} \sigma_\nu} {\rho} , \label{eqn:8}
\end{align}
where $n_{\rm H}$ is the number density of H atoms and $\rho$ is the gas density in the white dwarf atmosphere.

\subsection{Potential energy curves for H$-$H$_{2}$ collisions}
From \cite{Peng1995}, the four lowest potential energy surfaces of H$_3$ are labelled as $E_1$, $E_2$, $E_3$ and $E_4$. The ground state energy ($E_1$) relates to H$_2(X^1\Sigma_g)+$H($1s$) while the first excited state $E_2$ corresponds to H$_2 (b^3\Sigma_u)+$H($1s$). $E_2$ correlates with the unbound electronic state of the H$_2(b^3\Sigma_u)$ molecule rather than the excited state of the H-atom \citep{Rohrmann2011}. This transition is spin-forbidden at large H–H$_2$ separations, and while weakly allowed transitions can occur during close-range interactions due to spin-orbit and vibronic couplings (at a rate of $\approx$0.5\% compared to the transition to $E_3$; \citealt{Peng1995}), the $E_2$ surface becomes nearly degenerate with $E_1$ at equilateral triangular geometry of H$_3$ constituting a well-known Jahn-Teller system. Therefore, we do not consider the transition from $E_1$ to $E_2$.

The next two levels $E_3$ and $E_4$ correspond to H$_2$($X^1\Sigma_g)+$H($2s$) and H$_2$($X^1\Sigma_g)+$H($2p$), respectively. They have identical energies in all geometries and are the two lowest Rydberg states of H$_3$.

We used the \textit{ab initio} data from \cite{boothroyd1996} for the potential energies of $\rm H_{3}$, which were obtained from accurate quantum mechanical calculations. The authors employed high-level configuration interaction methods, specifically multi-reference configuration interaction (MRCI), together with large Gaussian basis sets to capture the electronic correlation across a broad grid of nuclear geometries. The resulting global potential energy surface (commonly referred to as BKMP2) remains a benchmark for H$_3$ studies. While more recent computational methods may allow further refinement, a reassessment of the quantum chemistry calculations is beyond the scope of this work.

Their \textit{ab initio} data include information on the ground state and four excited state potential energies of H$_3$. Note that the $E_3$ and $E_4$ energies are indistinguishable in the H$_3$ surface energy data. Hence, we used the ground state and second excited state energy data of H$_3$ defined as $E_1$ and $E_3$, respectively, for our analysis. The energies are provided in units of hartrees, and are converted to electron volts by applying an offset of $-0.174496$\,hartree.

H$_2$ atoms vibrating in the ground state spend most of their time at the equilibrium separation of $\approx$1.4\,a.u. At $\Teff\leq6000$\,K, the average vibration amplitude is small ($\leq 0.24$\,a.u.) due to smaller vibrational excitation. We selected \textit{ab initio} data with $r_{\rm H_{2}}$ lying between 1.3 and 1.5 a.u. for obtaining the potentials from the H$_3$ surface. We found that the potentials do not vary enough over this range to necessitate a more detailed account of molecule vibrations, as highlighted by KS06. 

\subsubsection*{Dependence on H-H$_{2}$ collision angle}
\begin{figure}
\centering
\includegraphics[width=\columnwidth]{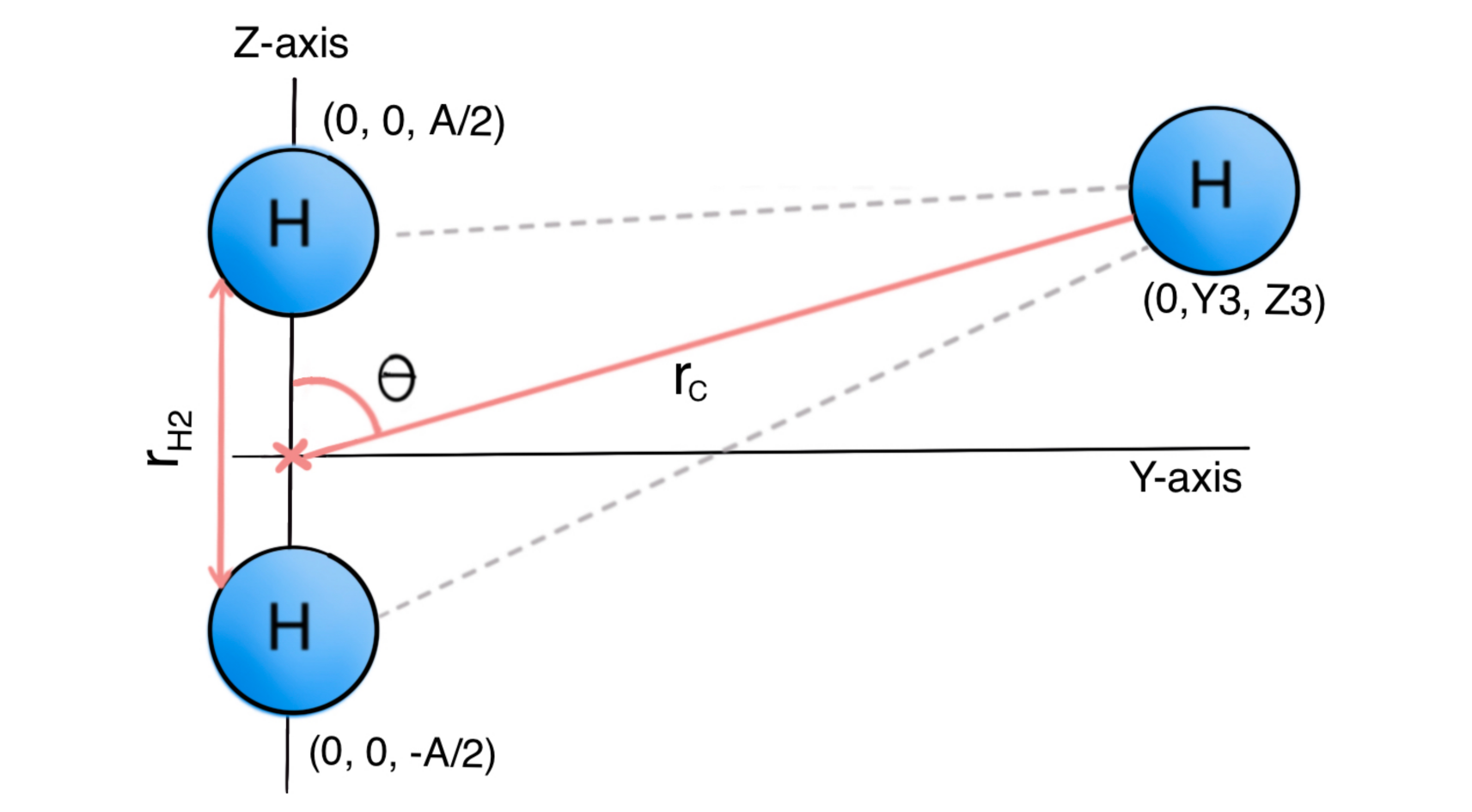}
\caption{Geometry for the interaction of the H$_{2}$ molecule with the H atom where $\theta$ is the collision angle, $|\vec{r}_{{\rm H}_{2}}|=A$ is the internuclear separation between the two H atoms in H$_{2}$, $\vec{r}_{c}$ is the distance between the centre of mass of H$_{2}$ to H atom. The positions of the three H atoms as defined in the centre of mass reference frame \citep{boothroyd1996} are labelled in the figure.}
\label{fig:H-H2geometry}
\end{figure}

KS06 have shown that the angular dependence of potential energies can be anisotropic for the H-H$_2$ collisions. Hence, we studied the effect of this dependence on the final opacities.
The geometry of the H$_{3}$ system is shown in Fig.\,\ref{fig:H-H2geometry}. Considering the center of mass of the  H$_{2}$ molecule, the collision angle $\theta$ between $\rm H_{2}$ and H is defined as
\begin{align}
    &{\rm cos}\,\theta= \frac {\vec{r}_{\rm H_{2}} \cdot \vec{r_{c}}}{|\vec{r}_{\rm H_{2}}| |\vec{r_{c}}|},
\end{align}
where $\vec{r}_{\rm H_{2}}$ is the separation between the two H atoms of the $\rm H_{2}$ molecule and $\vec{r_{c}}$ is the distance of the H atom from the centre of mass of the $\rm H_{2}$ molecule. The position information of the H atoms are taken from the \cite{boothroyd1996} data tables as labelled in Fig.\,\ref{fig:H-H2geometry} where $\vec{r}_{\rm H_{2}}=(0,0,A)$ and $\vec{r}_{c}=(0,Y_3, Z_3)$. Then, the collision angle is calculated using the above expression, where
\begin{align}
    &\theta={\rm cos^{-1}}\bigg(\frac{A\cdot Z_3} {A\cdot \sqrt{Y_3^2+Z_3^2}}\bigg).
\end{align}

\begin{figure*}
\centering
\includegraphics[width=0.33\textwidth]{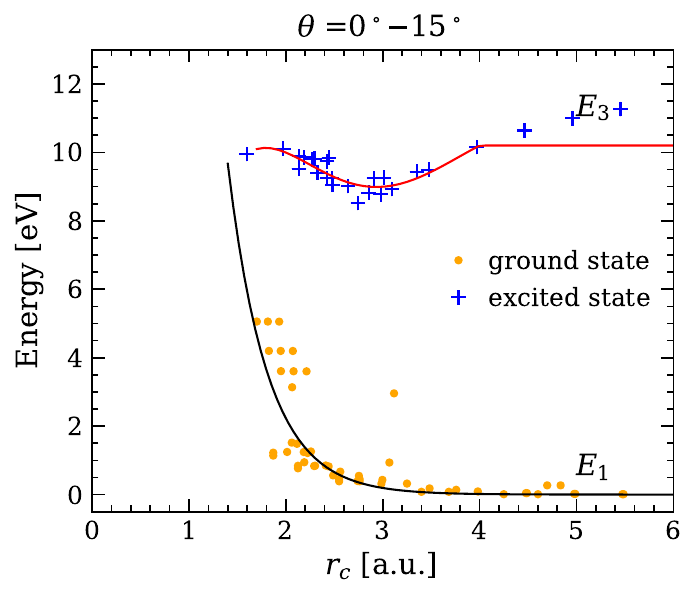}
\includegraphics[width=0.33\textwidth]{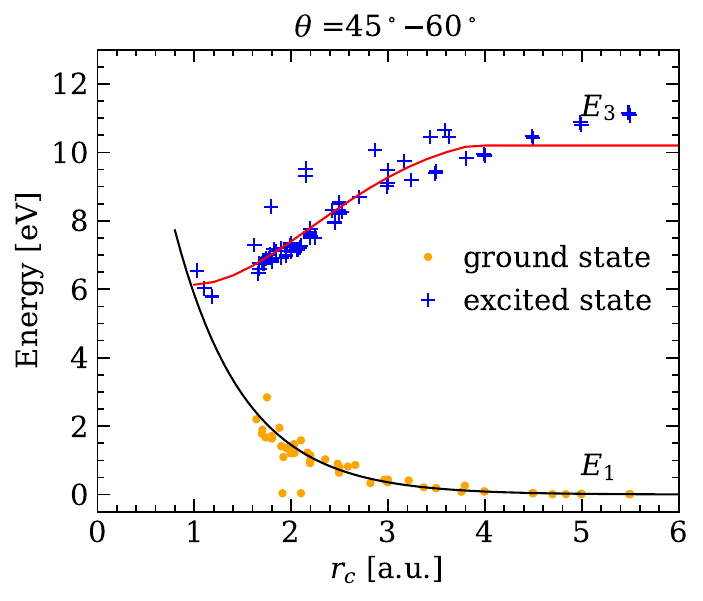}
\includegraphics[width=0.33\textwidth]{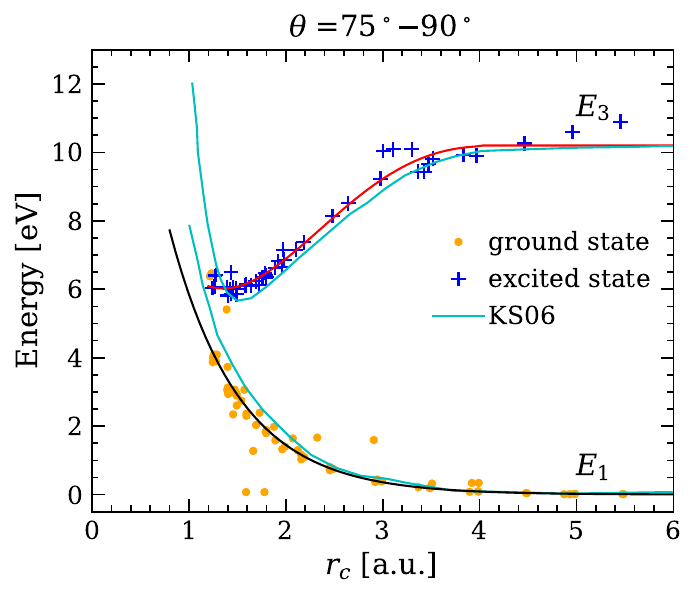}
\caption{H$-$H$_{2}$ potential energy curves for the ground state (black) and first excited Rydberg state (red) at three different collision angles. The orange dots and blue pluses denote the \textit{ab initio} data for the ground state and first excited state of the H$_{3}$ system \citep{boothroyd1996} where $\vec{r}_{{\rm H}_2}=1.40$ a.u. The solid black and red lines represent the polynomial fits to the ground and excited state energies, respectively. The potential energy curves used by KS06 for the collision angle of 90$^{\circ}$ are shown in solid cyan lines for comparison (right panel).}
\label{fig:potential_curve}
\end{figure*}

To study the angular dependence, we divided the potential energy data into six angular bins between 0 to 90$^\circ$ (bin width of 15$^\circ$) as shown in Fig.\,\ref{fig:potential_curve}. Since we have sparse data while considering angular binning for the ground state energy $E_{1}$, especially at $r_{c}\leq2$\,a.u., fitting an exponential functional form is difficult. Hence, we performed an exponential fit to the ground state energies considering two angular bins 0-30$^{\circ}$ and 30-90$^{\circ}$, to account for the anisotropy. An exponential fit to the energies overplotted onto the data of three angular bins is shown in Fig.\,\ref{fig:potential_curve} as an example. We find that the exponential curve is comparatively steeper for the lower angular bins, in agreement with the previous study (KS06). We notice a spread in the angle-dependent \textit{ab initio} data around the fit, especially for the low angular bins ($<60^\circ$). While its precise origin is not explicitly discussed in \cite{boothroyd1996}, this could arise partly due to variations of angle and $r_{\rm H_{2}}$. 

In the case of the lowest Rydberg state $E_3$, we have sufficient data spanning a larger range in $r_{c}$ for different $\theta$ bins. Therefore, we performed a polynomial fit (of higher orders 4-5) to $E_3$ potential energy curves at each angular bin. In the fit, we assumed that at larger separations $r_{c}>4$ a.u., the energy difference $E_3-E_1 \to$ 10.2\,eV corresponding to the Ly\,$\alpha$ transition energy for the isolated H atom at 1216\,\AA. Example fits of $E_3$ for three different angular bins are shown in Fig.\,\ref{fig:potential_curve}. The potential well has a larger depth for the angular bin 75$-$90$^\circ$ corresponding to a minimum energy of $\approx$6\,eV at equilibrium separation $r_c \approx 1.4$\,a.u., which is $\simeq$0.2\,eV higher in comparison to the energy curves used by KS06. The $E_3$ potential well gradually becomes shallower, shifting to larger equilibrium separations for lower angular bins (for instance, 3 a.u. for collision angles $<15^{\circ}$).  

\begin{figure}
\centering
\includegraphics[width=\columnwidth]{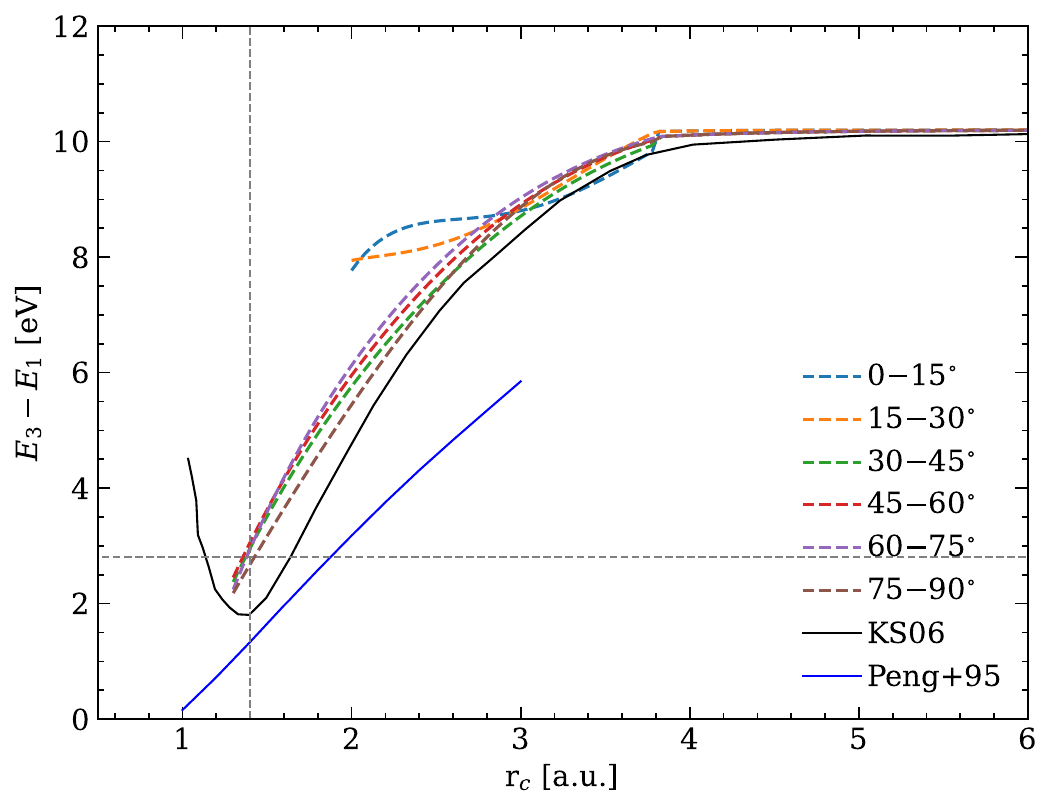}
\caption{The energy difference between the ground state ($E_1$) and the lowest Rydberg energy state ($E_3$) obtained from polynomial fits to the data for various collision angles. The potential energy difference from KS06 and \citet{Peng1995} for $D_{3h}$ geometry (equilateral triangle) is shown for comparison.}
\label{fig:energy_diff}
\end{figure}

We studied the energy difference $E_3-E_1$ for different collision angles obtained from the fits to the potential energies. From Fig.\,\ref{fig:energy_diff}, we note that the excitation energy is the highest (above 8 eV) for low collision angles $\leq30^{\circ}$. In contrast, the excitation energy is minimum for collision angles 75-90$^{\circ}$ with $E_3-E_1$ $\simeq2.8$\,eV at $r_c=1.4$\,a.u. This is roughly 1\,eV higher than KS06 due to differences in the energy curves (Fig.\,\ref{fig:potential_curve}). These differences become even larger compared to \cite{Peng1995} for the equilateral triangle configuration, where they found $E_3-E_1\simeq1.5$\,eV (0.3\,eV lower than KS06), based on different \textit{ab initio} data than \cite{boothroyd1996}. Overall, these comparisons highlight the complexities associated with the H-H$_2$ interaction energies. The effect of these uncertainties on the Ly\,$\alpha$ wing opacity is provided in the discussion.

\subsubsection*{Dipole moments}
We used the electric dipole transition moments from \cite{Peng1995} for the opacity calculations. According to \cite{Peng1995}, the interaction of H$_2$ with the H-atom leads to two lowest Rydberg states (defined as $E_3$ and $E_4$) having identical energies but significantly different transition dipole moments. However, this dipole moment vanishes for $E_4$ in the equilateral triangular configuration (defined as $D_{3h}$ geometry) corresponding to $\theta=90^{\circ}$ and $r_c=1.2$\,a.u. Since this is the critical region for optical opacity, we have only considered the contribution to the dipole moment from $E_3$ in the line profile computations. We fitted a cubic polynomial to determine the variation of dipole moment with separation $r_c$. The dipole moment is higher for separations $<2$\,a.u. and reaches 0.74 a.u. at larger separations as $r_c\to\infty$, which is the value of the electric dipole moment for an isolated H atom.  

\section{Results}\label{sec:res}
\subsection{Opacities}
\begin{figure}
\centering
\includegraphics[width=\columnwidth]{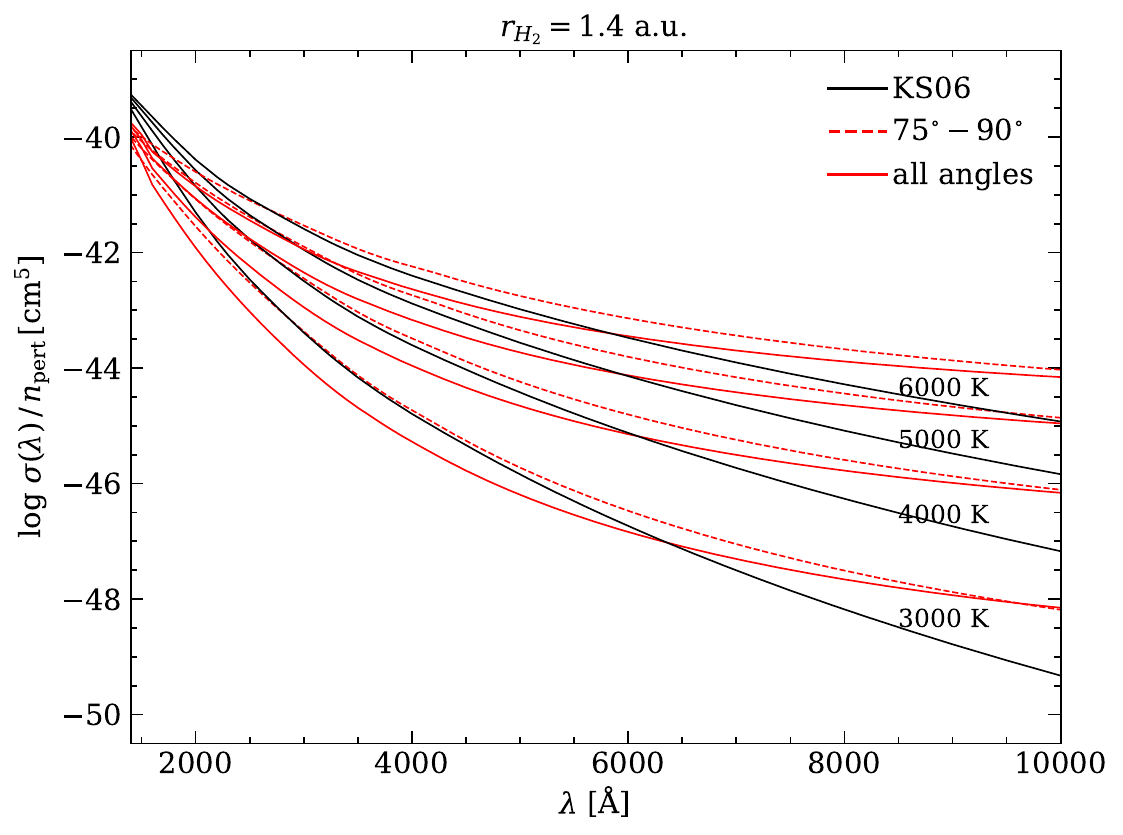}
\includegraphics[width=\columnwidth]{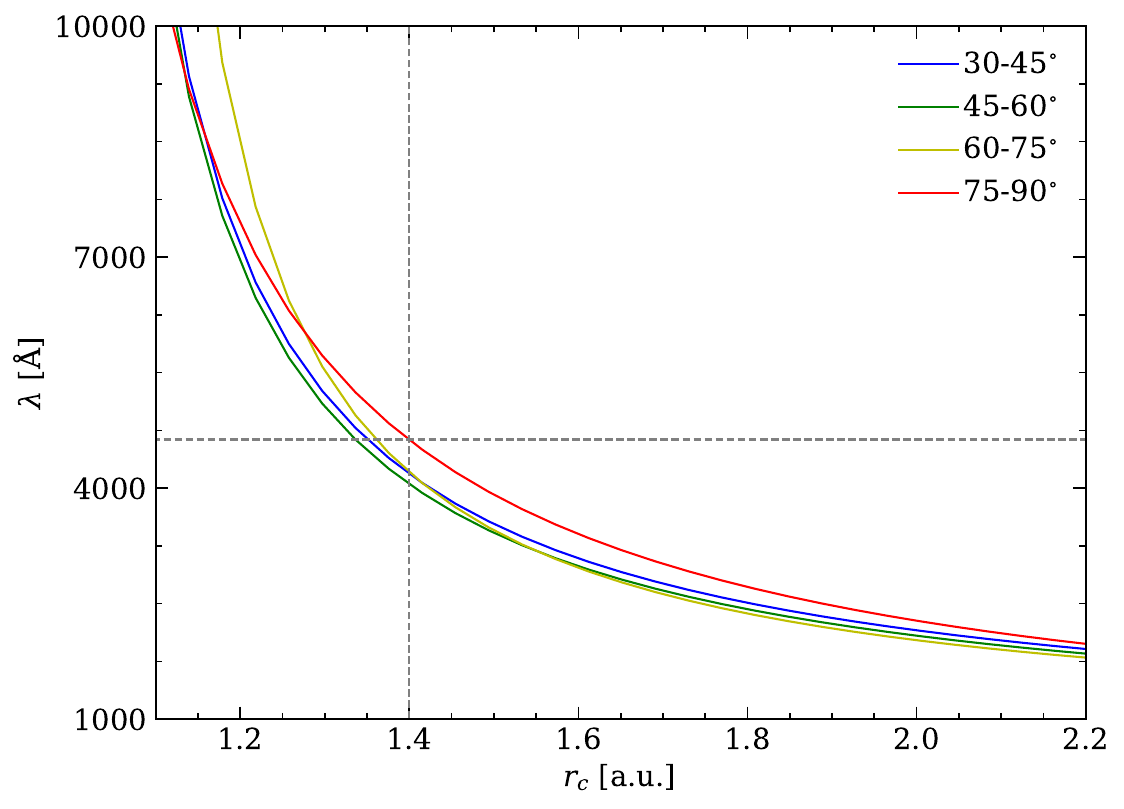}
\caption{\textit{Upper panel}: Absorption cross-section of the Ly\,$\alpha$ red wing per unit perturber density where the pressure broadening is caused by collisions of the H atom with H$_{2}$ for \Teff\ values of 3000, 4000, 5000, and 6000\,K. The dashed red lines are the values for $\theta$ between 75 to 90$^{\circ}$ while the solid red lines are for the case considering all H-H$_2$ collision angles. The opacity from KS06 \citep{kowalski2006} is shown in solid black lines for comparison. \textit{Lower panel}: Variation of the wavelength as a function of H-H$_2$ separation $r_c$, corresponding to the energy difference $E_3-E_1$ considering different collision angles. The dashed grey lines denote the wavelength 4600\,\AA\ corresponding to a separation of 1.4\,a.u. for collision angles 75-90$^{\circ}$.}
\label{fig:ly_opac}
\end{figure}

Using the fitted H-H$_2$ interaction energies and dipole moments, we calculated the absorption cross-section from Eqn.\,\ref{eqn:7} for various \Teff\ values taking into account the angle integral as shown in Fig.\,\ref{fig:ly_opac} (upper panel). We find that interactions at low collision angles $\theta\leq30^{\circ}$ lead to smaller absorption cross-sections due to their high excitation energies (Fig.\,\ref{fig:energy_diff}) and thus have a negligible contribution to the Ly\,$\alpha$ wing opacity. On the contrary, the interactions at collision angles 75$-$90$^\circ$ have the largest cross-section when compared with the contributions from all angles. Hence, they have a dominant effect on the wing opacity where a separation of $r_c=1.4$\,a.u. translates to a wavelength of $\approx4600$\,\AA\ as shown in Fig.~\ref{fig:ly_opac} (lower panel). The wavelength range spanning from 3000$-$7500\,\AA\ corresponding to H-H$_2$ separations of $1.2\leq r_{c}\leq1.7$\,a.u. plays a decisive role in defining the absorption in the optical spectrum, thus contributing to the wing opacity in cool white dwarf atmospheres. 

We also calculated the cross-sections considering only the angular bin 75-90$^\circ$ in order to compare with KS06 whose opacity is obtained from the energy levels with collision angle close to 90${^\circ}$. Both calculations agree at shorter wavelengths ($\lambda<5000$\,\AA) for the temperature range 3000$-$6000\,K while our opacity is $\approx$1\,dex larger at longer wavelengths close to 10\,000\,\AA. In contrast, when considering all angles, our cross-section is about 0.5\,dex smaller than KS06 for wavelengths shorter than 5000\,\AA.

The final opacity is derived from the absorption cross-section using Eqn.\,\ref{eqn:8}, with $n_{\rm H_2}$ and gas densities from white dwarf model atmospheres. 

\subsection{Multiple-perturber approximation}
As an additional test, we have relaxed the one-perturber approximation in the quasi-static opacity calculation. Using the same potential energy curves and dipole moments, we have used another code with a multi-perturber quasi-static approach following \citet{allard1991}. In this method, based on the unified line profile theory, instead of using the standard one-perturber autocorrelation function, based on a perturber passing the emitter with velocity $v$, we adopt the limit $v \rightarrow 0$ \citep[see eq.\,59 of][]{Allard1982}. This static-limit correlation function is then used analogously to the impact approximation, enabling the construction of a multi-perturber profile, which is subsequently Fourier transformed to give the line shape.

\begin{figure}
\centering
\includegraphics[width=\columnwidth]{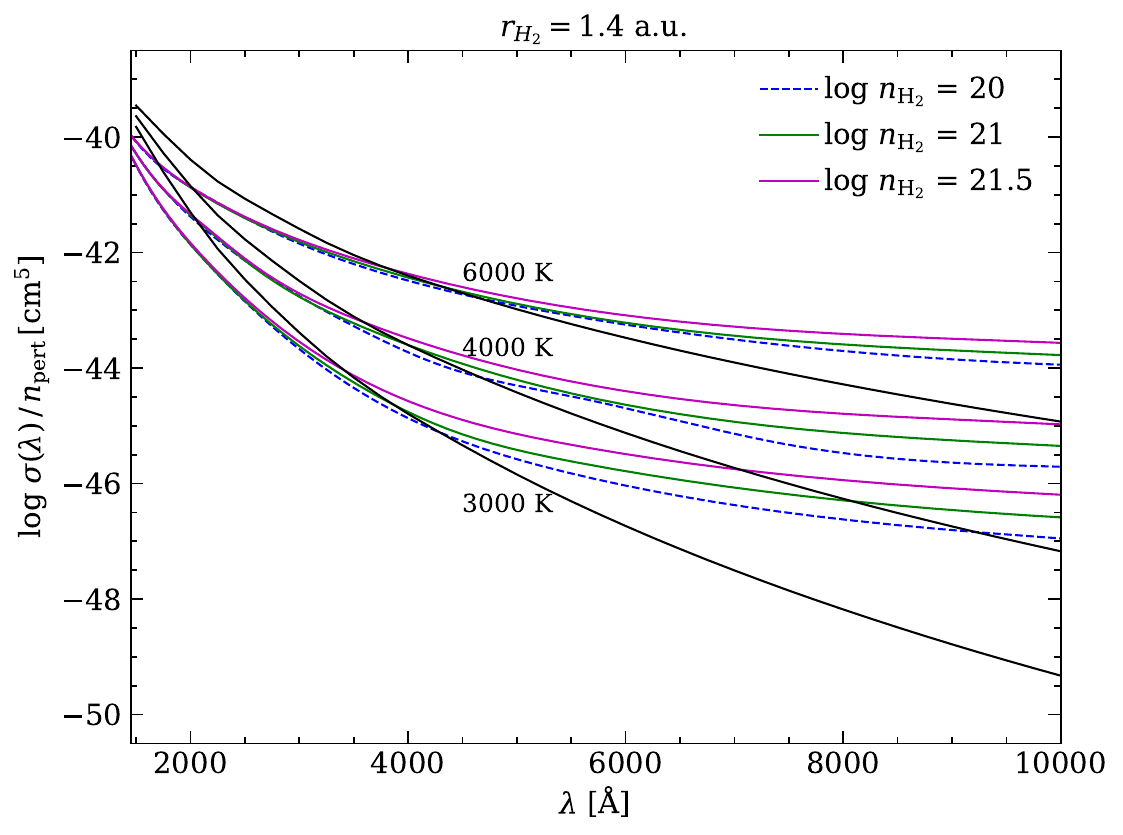}
\caption{Absorption cross-section of the Ly\,$\alpha$ red wing per unit perturber density where the pressure broadening is caused by collisions of the H atom with H$_{2}$ for \Teff\ values of 3000, 4000, and 6000\,K for different $n_{\rm H_2}$ number densities under the quasi-static multiple-perturber approximation. The opacity from KS06 \citep{kowalski2006} is shown in solid black lines for comparison.}
\label{fig:ly_opac_multip}
\end{figure}

The resulting quasi-static profile agrees closely with the unified theory profile \citep{Allard1982} in the line wings, if there are no extrema in the potential energy curves. The results of this calculation are shown in Fig.~\ref{fig:ly_opac_multip}. The opacity is very similar to that of the one-perturber quasi-static calculation at shorter wavelengths ($\lambda<5000$\,\AA), but increases significantly at longer wavelengths.

\subsection{Model atmospheres}
\begin{figure}
\centering
\includegraphics[width=\columnwidth]{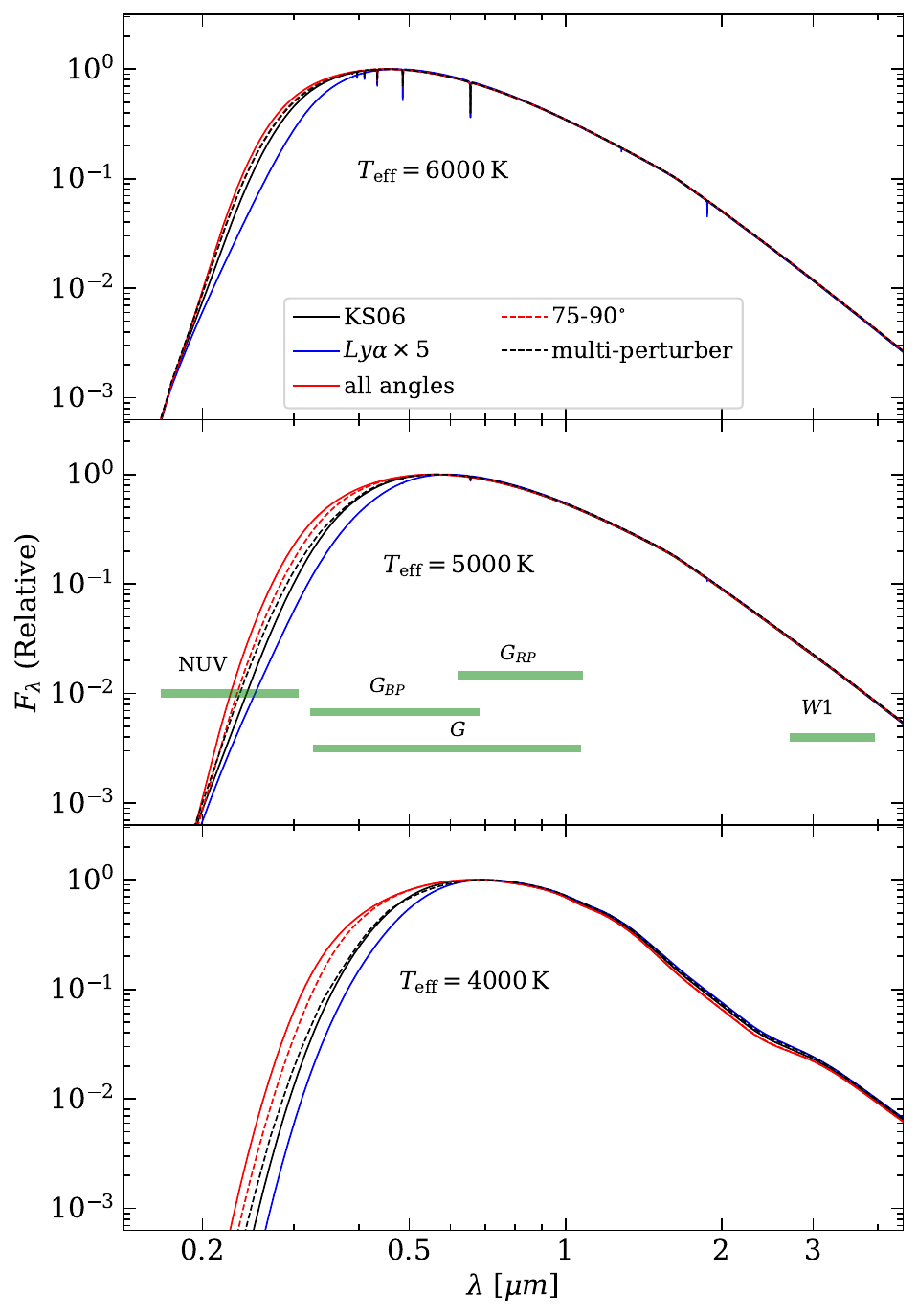}
\caption{Model spectra (solid red lines) calculated for pure-H atmospheres considering the new Ly\,$\alpha$ wing opacity from all H-H$_2$ collision angles in this work for \logg\ = 8.0 and \Teff\ = 4000\,K (lower panel), 5000\,K (middle), and 6000\,K (upper). The model spectra based on KS06 are shown in solid black lines for comparison with our spectra based on the wing opacity for collision angles 75-90$^\circ$ (dashed red). The model spectra for all angles using the quasi-static multi-perturber approximation are shown with dashed black lines. The model spectra (solid blue lines) considering five times the strength of the Ly\,$\alpha$ H-H$_2$ opacity from KS06 are also shown for comparison. The wavelength ranges covered by the \textit{GALEX} (\textit{NUV}), \textit{Gaia} ($G$, $G_{\rm BP}$ and $G_{\rm RP}$), and \textit{WISE} ($W1$) bandpasses are marked with thick solid green lines in the middle panel.}
\label{fig:mod_spec}
\end{figure}

To test the new opacities, we used the white dwarf model atmosphere code from \cite{Tremblay2011} that is currently based on the line profiles of \cite{Tremblay2009} with Ly\,$\alpha$ red wing opacities from KS06. Absolute fluxes are scaled using the mass-radius relation of \cite{Bedard2020}. The model grid covers \Teff\ ranging from 2000$-$40\,000\,K and \logg\ from 7.0$-$9.0\,dex. The model spectra and their comparison with KS06 for various \Teff\ and $\logg=8$ are shown in Fig.\,\ref{fig:mod_spec}.
The predicted fluxes are higher at shorter wavelengths $\lambda\leq$5000\,\AA\ when comparing models using the new opacity calculation with the standard ones based on KS06. In contrast, in the case of models based on the arbitrary assumption that the strength of the Ly\,$\alpha$ wing opacity is five times higher \citep{Brien2024}, we notice that the blue fluxes near 3000\,\AA\ are about 2--3 times lower than KS06, and 4--10 times lower than our new opacity calculation considering the angle integral, respectively. The flux differences increase with decreasing temperature, mainly from the increasing H$_2$ number density. In addition, the \textit{Gaia} ($G$ and $G_{\rm BP}$) and \textit{GALEX} (NUV) bandpasses fall in the wavelength range 1700$-$5000\,\AA, where we notice the maximum variation between the models. On the other hand, the fluxes are significantly less affected in the wavelength range 6300$-$10\,600\,\AA\ spanned by the \textit{Gaia} $G_{\rm RP}$ filter. 

\subsection{SEDs of cool white dwarfs}
\begin{figure}
\centering
\includegraphics[width=\columnwidth]{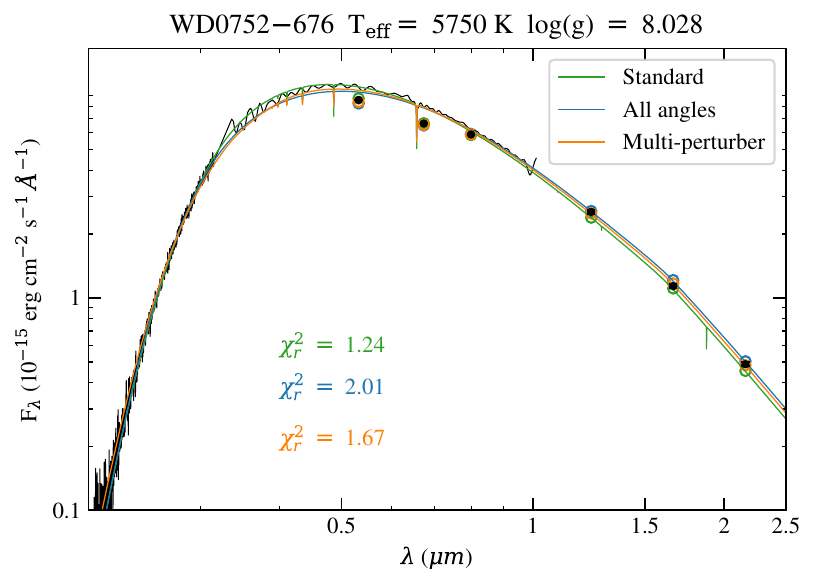}
\includegraphics[width=\columnwidth]{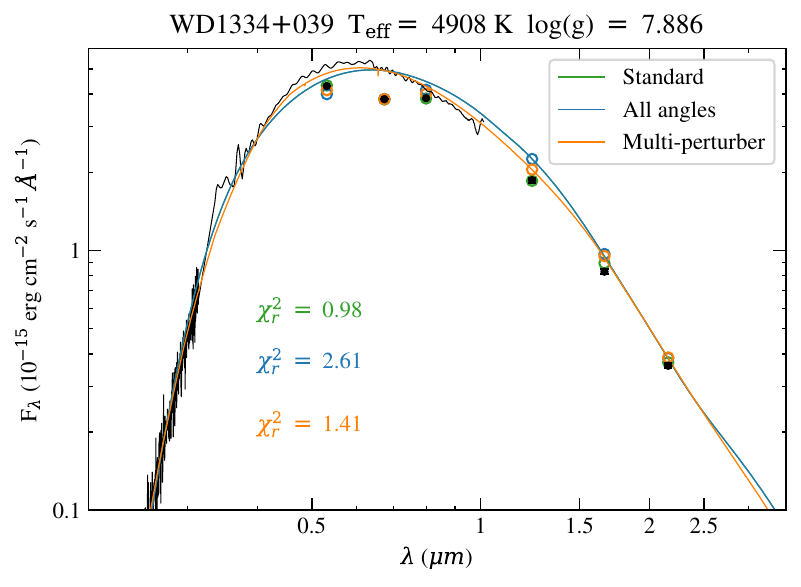}
\caption{SEDs of two cool DA white dwarfs with $\Teff\leq6000$\,K from the 13 pc sample constructed using \textit{HST} STIS spectroscopy and \textit{Gaia} XP spectra (solid black line), as well as \textit{Gaia} and 2MASS photometry (black dots). Model fits to the SEDs are shown for the Ly\,$\alpha$ H-H$_2$ opacity from KS06 (green), and for the updated opacities derived in this work, considering all H-H$_2$ collision angles under the quasi-static approximation (blue), and the multiple-perturber approximation (orange). The synthetic photometry for the three models is shown in open circles. The reduced $\chi^2_r$ values for the three model fits are labelled to illustrate the relative fit quality. The atmospheric parameters at the top are using the KS06 opacity.}
\label{fig:13pc_sed}
\end{figure}

We tested the different opacities for selected objects in a sample of cool white dwarfs within 13\,pc to assess their impact on the low mass problem. Spectral energy distributions (SEDs) were constructed for two cool DA white dwarfs that have \textit{HST} STIS spectroscopy (program 14076) by combining these data with \textit{Gaia} XP spectra, $G$, $G_{\rm BP}$, and $G_{\rm RP}$ photometry \citep{Gaia2023}, and 2MASS $JHK$ photometry \citep{2MASS2006}. Model fits to the SEDs of DA white dwarfs WD\,0752$-$676 (5750\,K, $\logg=8.03$) and WD\,1334$+$039 (4908\,K, $\logg=7.89$), using the standard (KS06) and updated opacities, both with the quasi-static and multiple-perturber approximations, are shown in Fig.\,\ref{fig:13pc_sed} as examples. To evaluate the quality of the fit, we computed the reduced $\chi^2$ ($\chi^2_r$) for each model opacity. We find that the parameters agree for all white dwarfs when considering only \textit{Gaia} and 2MASS photometry, irrespective of the adopted Ly\,$\alpha$ opacity. However, when including \textit{HST} STIS spectroscopy, the $\chi^2_r$ values are higher for the new opacity with the quasi-static approximation compared to KS06 models and those considering the multiple-perturber approximation. Overall, the SED fits with the updated opacities do not yield any visual improvement over the standard models.

\subsection{Application to the 40\,pc white dwarf sample}
\begin{figure*}
\centering
\includegraphics[width=\textwidth]{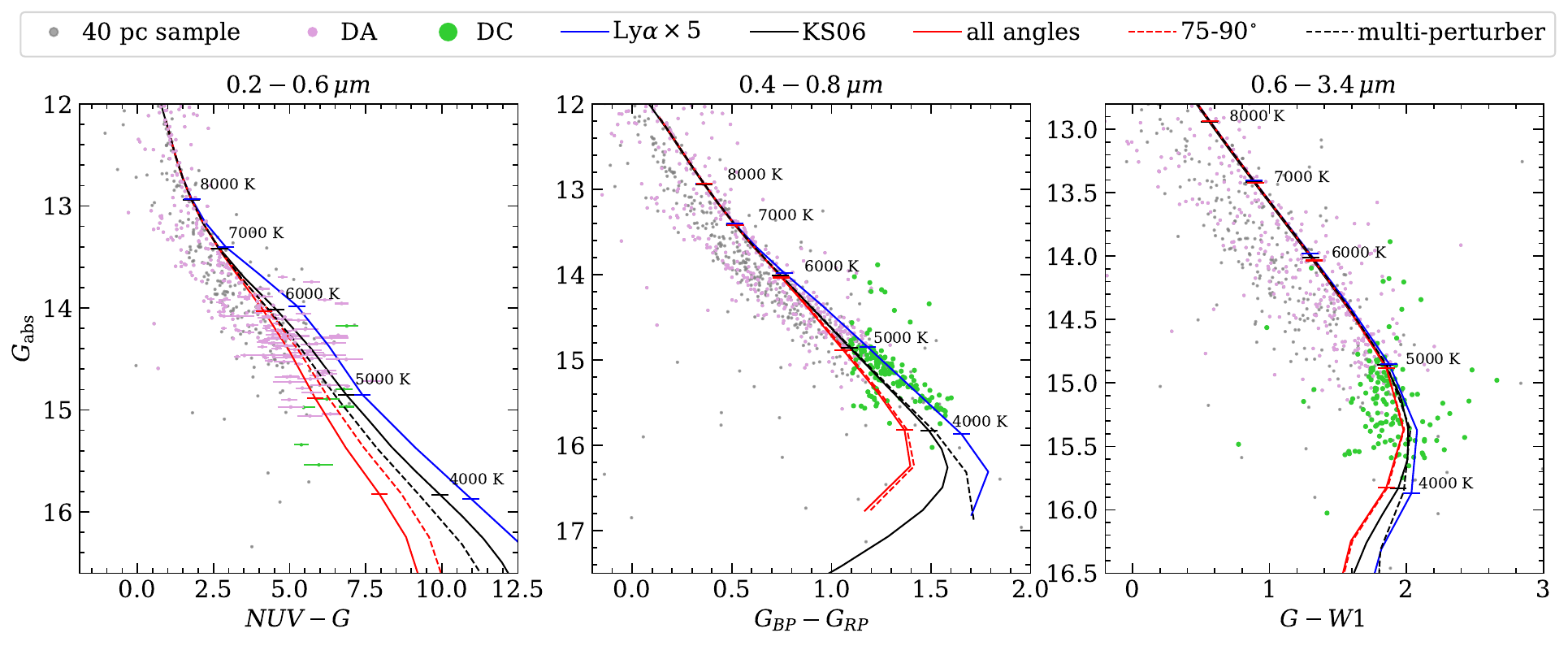}
\caption{H-R diagrams for the 40\,pc white dwarf sample spanning UV ($0.2\,\rm \mu m$) to infrared wavelength ($3.4\,\rm \mu m$) ranges \citep{Brien2024} where DA white dwarfs are highlighted in violet. Cool ($\Teff\leq5000$\,K) DC white dwarfs, suffering the most from the low mass-problem, are highlighted with green dots. The evolutionary tracks \citep{Tremblay2011,Bedard2020} for $\logg=8.0$ are based on the Ly\,$\alpha$ H-H$_2$ opacity from KS06 (black), Ly$\,\alpha\times$5 of KS06 from \citet{Brien2024} (blue) and the opacity derived in this work considering all H-H$_2$ collision angles (solid red), only angles 75-90$^{\circ}$ (dashed red) and the multiple-perturber approximation (dashed black). The short horizontal lines represent the \Teff\ from the model tracks to guide the eye. The model track (solid red) considering H$_3$ potential energies for all collision angles is still unsuccessful in reproducing the observed \textit{Gaia} colours of cool DC white dwarfs in $G_{\rm abs}$ vs $G_{\rm BP}-G_{\rm RP}$ (middle panel), however providing a better fit to the $NUV-G$ colours (right panel) for $\Teff < 7000\,$K.}
\label{fig:40pc_hrd_lya}
\end{figure*}

We also tested the different model opacities with the volume-limited sample of white dwarfs within 40 parsecs of the Sun \citep{Brien2024}. The sample consists of 1076 spectroscopically confirmed objects, among which 61.1\,per cent are of DA spectral type and 27.0\,per cent are of DC spectral type. Below 5000\,K where the low mass problem is most prominent, 69.7\,per cent are DC white dwarfs. 
Since it is difficult to distinguish between H- or He-rich atmospheres in cool DC white dwarfs, we compare our pure-H model spectra to the entire white dwarf population. Although a considerable fraction ($\approx$25\,per cent) of cool DC white dwarfs are expected to have He-rich atmospheres \citep{Elms2022,Brien2024}, both compositions empirically overlap in the near-UV, optical and near-IR HRDs, and therefore the distinction is not critical for our comparison with the synthetic magnitudes. In practice, using either the full DC population or only the subset of suspected H-atmosphere white dwarfs would yield similar HRDs to fit, differing primarily in the total number of objects included.
We cross-matched the 40\,pc sample with the \textit{GALEX} \citep{martin2005} and \textit{WISE} \citep{wright2010} photometric catalogues to construct multi-wavelength HRDs for colour combinations of \textit{Gaia} with \textit{GALEX} $NUV$ and \textit{WISE} $W1$ filters spanning the wavelength range 0.2$-$3.4$\,\mu$m.

Convolving the model spectra with the filter throughputs, we calculated the magnitudes in the \textit{GALEX}, \textit{Gaia}, and \textit{WISE} filters. To compare with the observed colours and magnitudes, we used cooling tracks at a fixed $\logg=8$, which corresponds to the median mass of $\approx$0.6\,\Msun\ observed for warmer DA white dwarfs \citep{Brien2024} and expected for cooler DC white dwarfs cooling at constant mass. The observed HRDs overplotted with cooling tracks for different Ly\,$\alpha$ red wing opacities are shown in Fig.\,\ref{fig:40pc_hrd_lya}. We note that all cooling tracks are coincident and reproduce the observations well at temperatures $>$6000\,K for the optical and IR, and $>$7000\,K for the UV. Below these \Teff, the standard KS06 track deviates from the observed distribution towards bluer $G_{\rm BP}-G_{\rm RP}$ colours (Fig.\,\ref{fig:40pc_hrd_lya}; middle panel), affecting especially the cool DC white dwarfs with $\Teff\leq5000$\,K ($G_{\rm abs}\geq$15\,mag). The implication of this offset is that white dwarfs have lower inferred masses and $T_{\rm eff}$ when fitting \textit{Gaia} photometry \citep{Brien2024}.
Considering the scenario where the KS06 Ly\,$\alpha$ opacity is arbitrarily multiplied by a factor of five \citep{Brien2024}, the models show a better fit to the $G_{\rm BP}-G_{\rm RP}$ colours, however, they deviate significantly when compared to the $NUV-G$ colours for $\Teff<7000$\,K ($G_{\rm abs}>13.5$\,mag). In contrast, model spectra with our new Ly\,$\alpha$ opacity show a better fit to the observed $NUV-G$ colours (Fig.\,\ref{fig:40pc_hrd_lya}; left panel) while disagreeing even more with the $G_{\rm BP}-G_{\rm RP}$ colours for DC white dwarfs. In the case of the infrared, there are only minor differences in the $G-W1$ colours between the models, with ours predicting bluer colours than others. This implies that the optical and UV parameters of cool white dwarfs are sensitive to the adopted Ly\,$\alpha$ broadening model, but the uncertainties in the angle interpolation of H$_{3}$ \textit{ab initio} potential curves, as well as the lack of multi-perturber collisions in previous calculations,  are clearly not the cause of the low-mass problem. 

\section{Discussion}\label{sec:discuss}
In our study, the potential energy curves utilised for opacity calculations are based on \textit{ab initio} data of H$_{3}$ from \cite{boothroyd1996} with full H-H$_2$ collision angle dependence, while KS06 used analytic BKMP2 surfaces based on the same data. A more recent H$_3$ energy surface has been presented by \citet{Mielke2002}. They report a maximum deviation of $\approx$0.25\,mH ($\approx$0.01 eV) compared to the energy surfaces from \cite{boothroyd1996}. We verified that the effect of these uncertainties on the opacity and final model colours is negligible.

In a previous study, \cite{Rohrmann2011} have also shown that the ground and excited state energies and therefore the opacities can differ depending on the nature of the data used for the H$_3$ potential energy surfaces. For example, they found that the data from \cite{KG1979} based on theoretical calculations at collision angle 90$^{\circ}$ are $\approx$ 0.8\,eV lower for distances $r_c <5$\,a.u. Indeed, using potential energies of \cite{KG1979} and \cite{RK1986}, they found that their absorption cross-sections decrease slightly at wavelengths larger than 4500\,\AA\ compared to KS06, due to a difference in potential energies at $r_c$ between 1.3 to 1.5\,a.u. However, since these changes were minor, they found a similar solution to KS06 when comparing the models with observed $B-V$ vs $V-K$ and $V-I$ colours. Having the advantage of a better precision with \textit{Gaia} photometry and astrometry for cool DC white dwarfs, we investigated the effect of using different potential energy data on the model colours and whether they can resolve the low-mass problem. We also compared the potential energies with studies from \cite{pet1988} and \cite{Peng1995} that report lower excitation energies than the current \textit{ab initio} data for different H-H$_2$ geometries. Specifically for equilateral $D_{3h}$ geometry, \cite{Peng1995} reported an excitation energy that is 1.5 eV lower at the equilibrium distance of 1.4 a.u. compared to \cite{boothroyd1996}. Lowering the excitation energy to their reported numbers, we find an improvement to the masses of cool DC white dwarfs ($+0.05\,\Msun$ compared to KS06 at 4000\,K); however it still fails to reproduce the observed $G_{\rm BP}-G_{\rm RP}$ colours and falls short of having cool white dwarfs evolving at a constant median mass of $\approx$0.6\,\Msun, as expected from single-stellar evolution. 

We also verified whether the dipole moments for different geometric configurations can alter the Ly\,$\alpha$ H-H$_2$ opacity and explain the discrepancy between models and observations. For this, we used the data from \cite{pet1988}, which provides dipole moments for collinear, insertion, and perpendicular paths of H-H$_2$ interactions. Considering these cases, we find that there are negligible changes in the model colours and masses when compared with \cite{Peng1995} for cool DC white dwarfs. 

\begin{figure}
\centering
\includegraphics[width=\columnwidth]{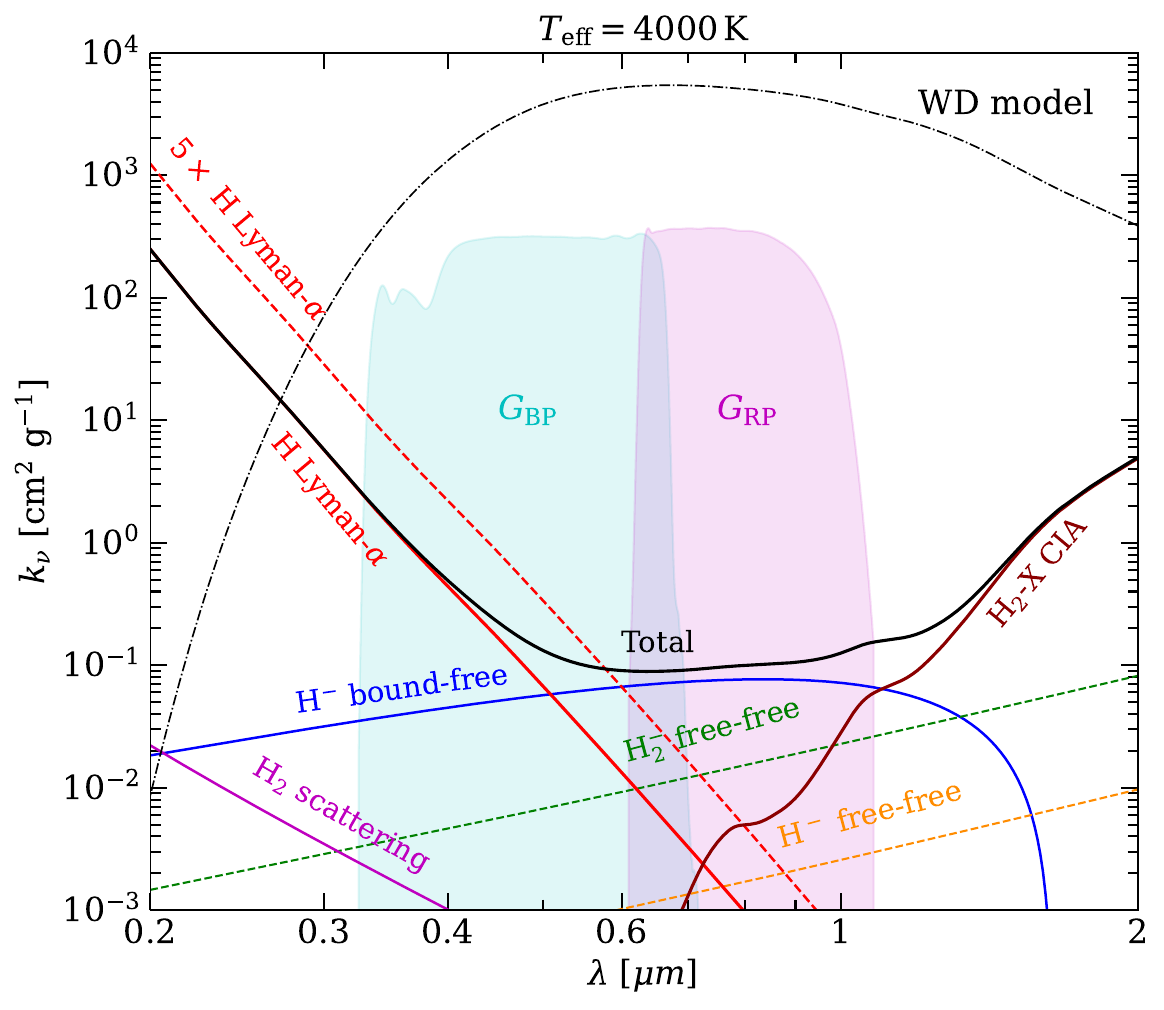}
\caption{Contributions of various opacities as a function of wavelength to the total radiative opacity (solid black) at the photosphere ($\tau_R=2/3$) of a cool H-atmosphere white dwarf with $\Teff=4000$\,K and $\logg=8$. The corresponding white dwarf synthetic spectrum (arbitrarily scaled in log F$_\lambda$) is shown as a dashed black line. In addition, we show as an illustration the Ly\,$\alpha$ $\times$ 5 ad-hoc opacity (dashed red) that effectively solves the low-mass problem in the optical. The normalised effective area curves of \textit{Gaia} filters $G_{\rm BP}$ and $G_{\rm RP}$  (in linear scale) are shown to highlight the dominant opacity sources in these wavelength regimes.}
\label{fig:all_opacity}
\end{figure}

In general, the above tests suggest that the opacity in the far red wing of Ly\,$\alpha$ arising due to H-H$_2$ collisions in the semi-classical quasi-static broadening approximation is unable to address the discrepancy between the model and \textit{Gaia} colours. Although the different variations of the new model opacity provide a better fit to the ultraviolet colours in the HRD, they result in low masses as determined from \textit{Gaia} photometry, similarly to previous studies. Therefore, we recommend the continued use of the ad-hoc mass and temperature corrections from \citet{Brien2024} applied to all white dwarfs cooler than 6000\,K, in the absence of a physical solution for the low-mass problem of cool DC and DA white dwarfs. 

\begin{figure*}
\centering
\includegraphics[width=\textwidth]{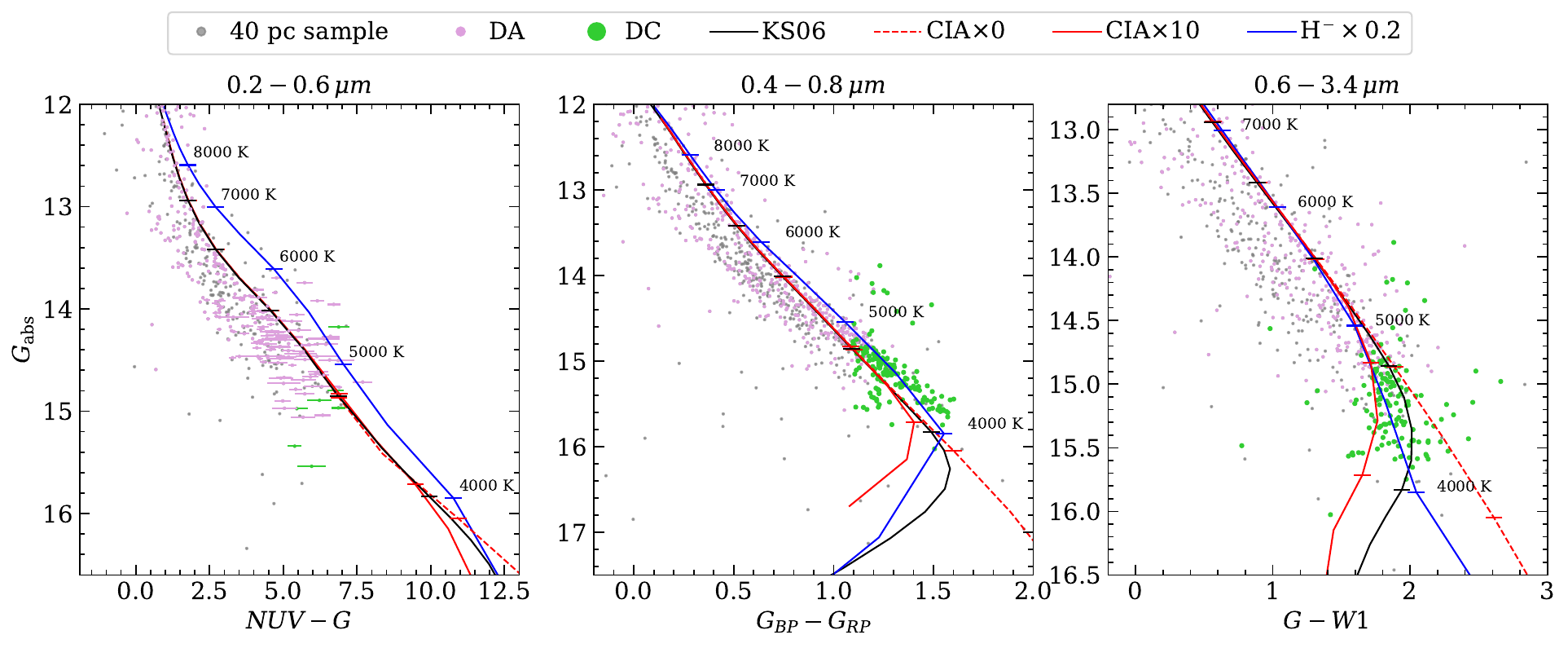}
\caption{Same as Fig.\,\ref{fig:40pc_hrd_lya} but the evolutionary tracks \citep{Tremblay2011,Bedard2020} for $\logg=8.0$ are based on the Ly\,$\alpha$ opacity from KS06 (black line), considering no CIA opacity, i.e. CIA$\times0$ (dashed red line), increasing the CIA H$_{2}$-H$_{2}$ opacity strength by an arbitrary factor of 10 (solid red line), and decreasing the bound-free H$^-$ opacity by a factor of 5 (blue line). The models with alternative H$^-$ opacity match relatively well the observed cooling track of the cool DC white dwarfs (green dots) in the near-infrared (left) and optical (middle), while deviating in the ultraviolet (right). On the contrary, the models with different CIA opacities are unable to match any of the observed distributions.}
\label{fig:40pc_hrd_cia_h-}
\end{figure*}

Apart from Ly\,$\alpha$ H-H$_2$ opacity, H$^-$ opacity arising due to bound-free absorption \citep{john1988} and CIA caused by the interactions of H$_2$-H$_2$ molecules \citep{borysow2001} are found to be the major sources of opacity in the optical and near-infrared ranges, respectively, as shown in Fig.\,\ref{fig:all_opacity} for a pure-H white dwarf photosphere at $\Teff=4000$\,K and $\log(g)=8.0$. We note that the Ly\,$\alpha$ wing opacity is the dominant opacity source in the wavelength range 2000$-$5000\,\AA, which is covered by the $Gaia$ $G_{\rm BP}$ filter. At longer wavelengths ($\gtrsim 5000$\,\AA), the H$^-$ bound-free opacity overtakes Ly\,$\alpha$, contributing most of the continuum opacity up to about 1\,\micron, spanned by the $G_{\rm RP}$ filter. Beyond this point, CIA becomes the primary opacity source. In Fig\,\ref{fig:all_opacity}, we illustrate that multiplying the Ly\,$\alpha$ opacity by a factor of five, an ad-hoc adjustment to address the low-mass problem \citep{Brien2024}, directly alters its balance with the H$^-$ bound-free opacity in the $Gaia$ $G_{\rm BP}$ and $G_{\rm RP}$ filters, respectively, resulting in redder colours. Motivated by this interplay, we varied the relative contributions of CIA and H$^-$ opacity in our models to assess their impact on the emergent spectra of cool H-atmosphere white dwarfs.

The CIA group of opacities is known to be highly uncertain \citep{Bergeron2022, Elms2022,blouin2024}, although many past studies are related to the CIA H$_2$-He opacity, which is seen at higher temperatures in denser He-rich atmospheres. To verify the role of the CIA opacity, we generated model fluxes by rescaling the strength of the CIA H$_2$-H$_2$ opacity from zero to ten times the standard model \citep{borysow2001}. We find that varying the CIA strength by these extreme amounts leads to a larger range of model colours in the IR while having a negligible effect in the NUV. For example, in the case of $G$ vs $G-W1$ HRD as shown in Fig.\,\ref{fig:40pc_hrd_cia_h-} (right panel), rescaling CIA by ten times shifts the model fluxes to shorter wavelengths, thus leading to bluer colours. However, the models are unable to fit the observed magnitudes and colours in the UV and optical. Hence, issues with CIA opacity are unlikely to be the explanation for the low-mass problem of cool DC white dwarfs. 

For clarifying the role of the H$^-$ bound-free opacity, we generated model spectra by arbitrarily decreasing the strength of the bound-free H$^-$ opacity by five times relative to the one currently used in the models \citep{john1988}. This opacity arises from a simple quantum system that is not generally considered problematic, and has been extensively and accurately studied since the early works of \cite{Chandra1945, Chandra1958} and \cite{Doughty1966}. However, H$^-$ has a low ionisation potential of 0.754\,eV, which means it is particularly sensitive to non-ideal gas effects in the denser atmospheres of cool white dwarfs.

Contrary to CIA opacity, we note that the rescaled H$^-$ opacity better explains the observed distribution of DC white dwarfs in $G$ vs $G_{\rm BP}-G_{\rm RP}$ (middle panel) and $G-W1$ HRD (right panel) as shown in Fig.\,\ref{fig:40pc_hrd_cia_h-}. However, it deviates by a large factor in $G$ vs $NUV-G$ HRD (Fig.\,\ref{fig:40pc_hrd_cia_h-}; left panel) from the observed distribution, with the rescaled models predicting brighter $G$ magnitudes and redder colours compared to the existing ones. 

\section{Conclusion}
We revisited the model atmospheres using the Ly\,$\alpha$ H-H$_2$ opacity of \citet{kowalski2006}  to investigate its role in the low-mass problem in cool white dwarfs ($\Teff\leq6000$\,K) based on \textit{Gaia} photometric observations. In this work, we explicitly utilised the \textit{ab initio} data from \cite{boothroyd1996} for fitting the H-H$_2$ interaction energies at various collision angles, and examined their influence in the final opacity and model calculations. Based on our new opacity calculation, we find that the model spectra are more consistent with the observed near-UV \textit{GALEX} colours of cool DA and DC white dwarfs within 40\,pc. However, detailed SED analyses of selected nearby objects with \textit{HST} STIS spectroscopy show no improvement over the standard models in the UV. Furthermore, the low-mass problem in optical photometry remains unresolved, and its origin is still largely unknown. Based on various opacity tests, we suggest that an improved model ($\Teff\leq6000$\,K) with either higher opacity in the wavelength range 0.2$-$0.5\,$\mu$m ($G_{\rm BP}$ filter) or a lower opacity in the wavelength range 0.5$-$1\,$\mu$m ($G_{\rm RP}$ filter) than the current standard models can provide a better solution to the low mass problem. Until then, applying an ad-hoc mass correction for white dwarfs with effective temperatures below 6000\,K when deriving their parameters from optical photometry appears to be the most practical approach for now \citep{Brien2024}. 

Currently, we detect only seven DC white dwarfs with $\Teff\leq5000$\,K at the sensitivity of \textit{GALEX} due to their low near-UV fluxes. This underscores the need for future UV telescopes (such as \textit{UVEX}; \citealt[]{Kulkarni2021}, \textit{CASTOR}; \citealt[]{cote2019}, \textit{INSIST}; \citealt[]{subr2022}) with better sensitivity and reaching fainter magnitudes ($G_{\rm abs}\geq15$\,mag) in order to validate the theoretical predictions with observations. Since they are relatively bright in the infrared, \textit{JWST} observations of a larger sample will also be encouraging to assess the role of H$^-$ and CIA opacities in the measurements of masses \citep{blouin2024}. 

\section{Acknowledgements}
We thank the anonymous referee for helpful comments that improved the manuscript. This research is based on observations made with the NASA/ESA Hubble Space Telescope obtained from the Space Telescope Science Institute, which is operated by the Association of Universities for Research in Astronomy, Inc., under NASA contract NAS 5–26555. These observations are associated with program 14076. SS, PET, MOB, and BTG received funding from the European Research Council under the European Union’s Horizon 2020 research and innovation programme number 101002408 (MOS100PC) and 101020057 (WDPLANETS). SB acknowledges the support of the Canadian Space Agency (CSA; 23JWGO2A10). VF was a summer student under the Warwick URSS scheme. 

\section{Data availability}
The \textit{ab initio} data of H$_3$ potential energy surfaces used for our analysis are available in \cite{boothroyd1996}\footnote{https://www.cita.utoronto.ca/~boothroy/bkmp2.html}. The STIS spectroscopy data underlying this paper are available in the
raw form via the \textit{HST} MAST archive under the program 14076.

\bibliographystyle{mnras}
\bibliography{ref}
\end{document}